\documentclass[aps,prl,twocolumn,superscriptaddress, 10pt]{revtex4-1}
\usepackage{graphics,graphicx, soul, amsmath}
\usepackage{qcircuit}
\usepackage{color}
\DeclareMathOperator{\Tr}{Tr}

\begin{document}


\title{Measuring the Renyi entropy of a two-site Fermi-Hubbard model on a trapped ion quantum computer}


\author{N. M. Linke}
\affiliation{Joint Quantum Institute, Department of Physics, and Joint Center for Quantum Information and Computer Science, University of Maryland, College Park, MD 20742, USA}
\email{linke@umd.edu}
\homepage{http://iontrap.umd.edu}
\author{S. Johri}
\affiliation{Intel Labs, Intel Corporation, Hillsboro, OR 97124}
\author{C. Figgatt}
\affiliation{Joint Quantum Institute, Department of Physics, and Joint Center for Quantum Information and Computer Science, University of Maryland, College Park, MD 20742, USA}
\author{K. A. Landsman}
\affiliation{Joint Quantum Institute, Department of Physics, and Joint Center for Quantum Information and Computer Science, University of Maryland, College Park, MD 20742, USA}
\author{A. Y. Matsuura}
\affiliation{Intel Labs, Intel Corporation, Hillsboro, OR 97124}
\author{C. Monroe}
\affiliation{Joint Quantum Institute, Department of Physics, and Joint Center for Quantum Information and Computer Science, University of Maryland, College Park, MD 20742, USA}
\affiliation{IonQ Inc., College Park, MD 20742, USA}


\date{\today}

\begin{abstract}
The efficient simulation of correlated quantum systems is the most promising near-term application of quantum computers. Here, we present a measurement of the second Renyi entropy of the ground state of the two-site Fermi-Hubbard model on a $5$-qubit programmable quantum computer based on trapped ions. Our work illustrates the extraction of a non-linear characteristic of a quantum state using a controlled-swap gate acting on two copies of the state. This scalable measurement of entanglement on a universal quantum computer will, with more qubits, provide insights into many-body quantum systems that are impossible to simulate on classical computers.
\end{abstract}


\maketitle

One of the striking differences between classical and quantum systems is the phenomenon of entanglement. Analyzing large entangled states is of considerable interest for quantum computing applications. This is particularly relevant to quantum chemistry and materials science simulations involving interacting fermions \cite{Aspuru05, Bravyi2002}, small versions of which have been simulated on few-qubit quantum computers \cite{OMalley2016, Lamata2015, Kandala2017}. 
Recently, a quantum algorithm was developed to construct the entanglement spectrum of an arbitrary wave function prepared on a quantum computer via measurement of the Renyi entropies \cite{Johri17}. In this Letter we measure the second Renyi entropy in a $5$-qubit circuit by implementing a controlled-swap (C-Swap) gate, and mitigate experimental errors by exploiting the symmetry properties of this gate. We note that previous measurements of the Renyi entropy such as \cite{Islam2015} were not implemented on universal machines and may not be easily generalizable to arbitrary Hamiltonians or scalable to larger systems. 

For a many-body quantum system ideally described by the state $|\Psi\rangle$ and composed of two subsystems $A$ and $B$, the $n$th Renyi entropy is given by $S_n=\frac{1}{1-n}\log(R_n)$, where
\begin{equation}
R_{n}=\Tr(\rho^n_A)
\end{equation}
is the trace of the $n$th power of the reduced density matrix $\rho_A=\Tr_B(|\Psi\rangle\langle\Psi|)$. For non-zero entanglement we have $R_2<1$, which has the same universality properties as the von Neumann entropy $S=-\Tr(\rho_A\log(\rho_A))$. Both are measures of the entanglement between $A$ and $B$, and provide valuable information about the underlying physics of the system. For example, the Renyi entropy can be used to distinguish many-body localized states from thermalized states \cite{Bauer2013,Iyer2013,Bardarson2012,Vosk2013,Burrell2007} through their time dependence and dimensional scaling law \cite{Eisert2010}, and to study topological order \cite{Kitaev2006,Levin2006} and quantum critical systems \cite{Vidal2003}.

The system under investigation for this work is the two-site Fermi-Hubbard model, which describes interacting electrons on a lattice \cite{Hubbard1963,Hubbard1964}. Despite its simplicity, it has been postulated as a model for complex phenomena such as high-temperature superconductivity. Since its behavior in the thermodynamic limit remains inaccessible to classical numerical techniques, it has become a prime candidate for simulation by quantum computers \cite{Wecker1_2015,Wecker2_2015}. 

\begin{figure}[t]
\includegraphics[width=\columnwidth]{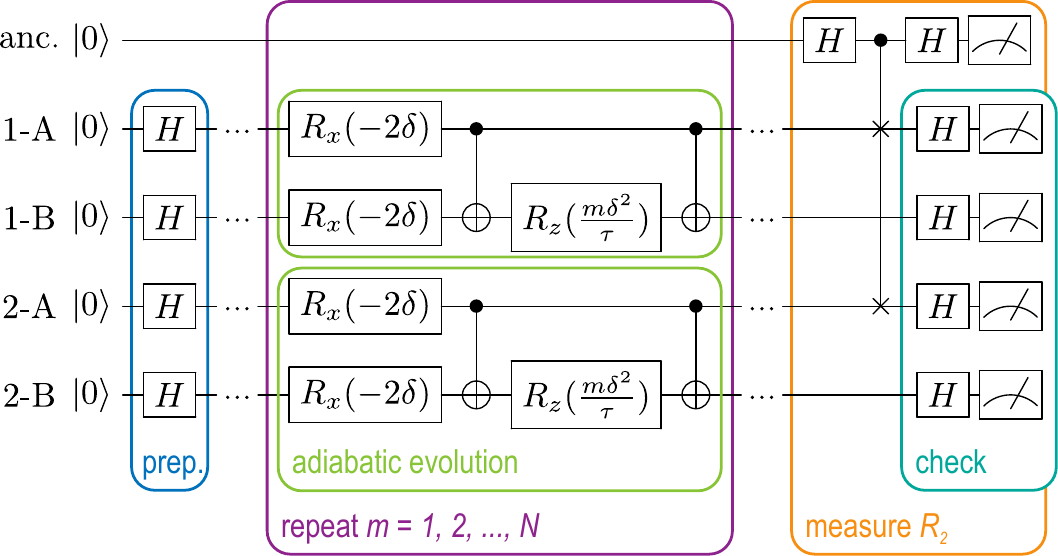}
\caption{The quantum circuit for the adiabatic evolution of two copies (labeled 1 and 2) of the two-site Fermi Hubbard model (each realized in two qubits $A$ and $B$), and the measurement of the second Renyi entropy. The non-interacting ground state is prepared via the application of Hadamard gates, followed by the digitized adiabatic evolution to a finite value of $U$ by repeated application of the central code block. $m$ is an integer referring to the $m$-th step of the adiabatic evolution; $\delta$ and $\tau$ are given in the text. The angles in the rotation gates are in radians. The Renyi entropy $R_2$ is measured by applying a C-Swap gate controlled by an ancilla qubit (anc.) on subsystems $A$ and subsequent detection of the state of the ancilla. By applying additional Hadamard gates and measuring the system qubits, experimental errors can be detected and discarded.}
\label{fig:circuit}
\end{figure}

\begin{figure*}[t]
\includegraphics[width=0.98\textwidth]{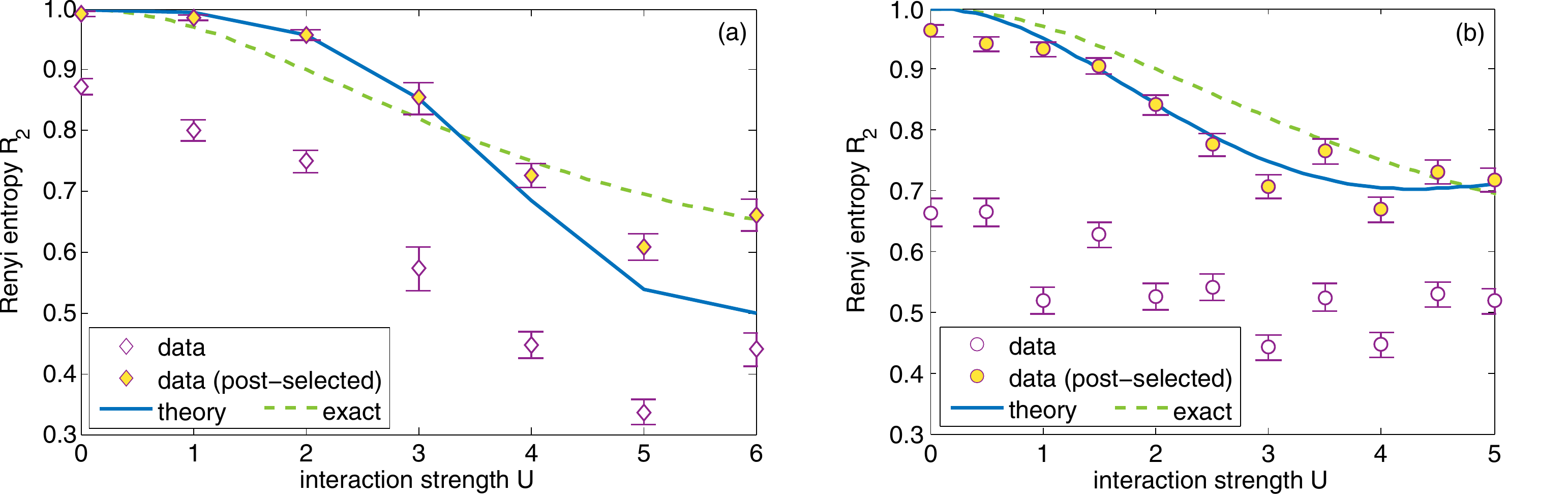}
\caption{Results of the measurement of $R_2$ after the digitized adiabatic evolution to different interaction strengths $U$ according to method I (a) and II (b) compared to the expected curve (solid line) and the exact solution (dashed line). The open symbols show the original data while the filled symbols show the same data after post-selection based on the symmetry of the C-Swap gate.}
\label{fig:compresults}
\end{figure*}

Our work consists of several co-designed theoretical and experimental steps. First, we find an efficient mapping from the electronic problem to the qubit space. Second, we develop a circuit for digitized adiabatic evolution to prepare the ground state of the model, parametrized by the Trotter step size and the total evolution time.  Based on available experimental resources and estimated Trotter errors, we choose a set of parameters that best corresponds to the result from exact diagonalization. Third, we realize the C-Swap gate, which is the key to efficiently extracting the second Renyi entropy. Finally, we integrate all of these elements into one circuit (fig. \ref{fig:circuit}) that prepares and evolves two copies of the Fermi-Hubbard system, and measures the second Renyi entropy with the help of an ancilla qubit. Importantly, in this step, we use the symmetry properties of the C-Swap gate to post-process the information contained in the four data qubits and reduce experimental errors. The results of this measurement for two sets of parameters are shown in Fig. \ref{fig:compresults}.

We execute the relevant quantum circuits on a trapped-ion system, that constitutes a programmable five-qubit quantum computer \cite{Debnath16} with full connectivity and an expressive native gate library \cite{LinkePNAS2017}. Qubits are realized in the hyperfine-split ground level of $^{171}$Yb$^+$ ions confined in a Paul trap (see Supplementary Materials). Single- and two-qubit gate fidelities are typically $99.1(5)\%$ and $98.5(5)\%$, respectively. Typical gate times are $10\:\mu$s for single- and $210\:\mu$s for two-qubit gates. The computational gates such as H, CNOT, and C-Swap are generated in a modular fashion by a compiler which breaks them down into constituent physical-level single- and two-qubit gates from the library (see Supplementary Material). 

The Fermi-Hubbard Hamiltonian is
\begin{align}
H=-t\sum _{\langle i,j\rangle ,\sigma }(c_{{i,\sigma }}^{{\dagger }}c_{{j,\sigma }}+c_{{j,\sigma }}^{\dagger }c_{{i,\sigma }})+U\sum _{{i=1}}^{N}n_{{i\uparrow }}n_{{i\downarrow }}
\end{align}
where $c_{{i,\sigma }}^{\dagger }$ and $c_{{i,\sigma }}$ are the electron creation and annihilation operators, respectively, for spin state $\sigma \in \{\downarrow, \uparrow\}$ and site $i$, and $n_{i,\sigma}=c_{{i,\sigma }}^{\dagger }c_{{i,\sigma }}$ is the electron number operator. Here $t$ is the hopping strength and $U$ is the on-site interaction. We consider the smallest non-trivial version of the model, which involves two sites and two electrons with total spin $S_z=0$ along the $z$-axis.

Typically, when mapping electronic problems to qubits, the Jordan-Wigner \cite{Jordan1928} or Bravyi-Kitaev \cite{Bravyi2002} mappings are used, both of which work in the second quantized basis. Here, the number of qubits $N_q$ is equal to the number of single-electron states $N_s$. Therefore, using the second quantization mapping would require $N_q=N_s=4$.

However, in many problems, there are additional symmetries or conservation laws that can reduce the size of the Hilbert space. For instance, the Hamiltonian above conserves both the number of electrons $N_e$ and the total spin along $z$, $S_z$. Therefore, a first quantization mapping, in which the size of the Hilbert space of the qubit system is equal to the size of the Hilbert space of the many-electron problem, makes the most efficient use of qubits. This is an important optimization for near-term quantum hardware, where the number of qubits available is limited.

In first quantization, the Hilbert space size is $4$, which can be mapped to two qubits as $|00\rangle=\{1_{\uparrow}1_{\downarrow}\}$, $|01\rangle=\{1_{\uparrow}2_{\downarrow}\}$, $|10\rangle=\{2_{\uparrow}1_{\downarrow}\}$, and $|11\rangle=\{2_{\uparrow}2_{\downarrow}\}$.  $\{i_{\sigma_i}j_{\sigma_j}\}$ represent the Slater determinants, which satisfy the number and spin conservation laws of the Hamiltonian. In this mapping, one qubit represents the up spin space and the other the down spin space. In this basis, the Hamiltonian is 
\begin{align}
H=
\begin{bmatrix} 
	U & -t & -t & 0 \\
	-t & 0 & 0 & -t \\
	-t & 0 & 0 & -t \\
	0 & -t & -t & U 
\end{bmatrix}.
\end{align}
This results in the qubit Hamiltonian (up to a constant, and scaling energy by $t$) \cite{Moll2016}
\begin{align}\label{Hamiltonian}
H =-(X_1+X_2)+\frac{U}{2}Z_1Z_2,
\end{align} 
where $X_i$ and $Z_i$ are Pauli matrices.

To prepare the ground state at finite $U$, we use digitized adiabatic evolution from the zero-interaction ground state. The time-dependent Hamiltonian is
\begin{align}
H(s)=-(X_1+X_2)+\frac{s}{2\tau}Z_1Z_2
\end{align}
from $s=0$ to $s=U\tau$ by linear interpolation. At $U=0$, the ground state is $(|0\rangle+|1\rangle)\otimes(|0\rangle+|1\rangle)$, which can be prepared with Hadamard gates (see Fig. \ref{fig:circuit}). 

The first order Trotterization for a Hamiltonian with two non-commuting terms $H_{a,b}$ is $\exp(-i(H_a+H_b)\delta)=\exp(-iH_a\delta)\exp(-iH_b\delta)+\mathcal{O}(\delta^2)$. For larger evolution time $\tau$ and smaller step size $\delta$, the approximation of the ground state is more accurate (see Supplementary Material). Here, at time $s=m\delta$, we set $H_a=-(X_1+X_2)$, $H_b=\frac{m\delta}{2\tau}Z_1Z_2$. Putting together Trotterization and digitized adiabatic evolution, we obtain the following sequence of unitary operations to prepare the ground state:
\begin{align}
V=\prod_{m=1}^{M}\bigg[\exp(i\delta X_1)\exp(i\delta X_2)\exp\bigg(-im\frac{\delta^2}{2\tau} Z_1Z_2\bigg)\bigg].
\end{align} Following \cite{Nielsen11}, we use the convention that $R_x(\theta)=\exp(-i\theta X/2)$, and similarly for $Z$. This leads us to the circuit for the digitized adiabatic evolution as shown in figure \ref{fig:circuit}.


We simulate the evolution on a classical computer to investigate the scaling of the error in $R_2$ in the first order Trotter approximation (see Supplementary Material). 
For the experimental implementation, there is a maximum number of gates that can be performed within the memory depth of the controller, which limits the length of a gate sequence to $8\:$ms. Since each Trotter step in the adiabatic evolution has a fixed number of gates, we bound the number of Trotter steps to $N_{\text{steps}} \leq 6$. We implement two different methods for evolving to finite $U$. 

Method I: For a fixed $\tau$ and $\delta$, one can sample at intervals of $\tau/\delta$ and $N_{\text{steps}}=U\tau/\delta$.  In order to evolve from $U=0$ to $U=6$, we choose $\tau/\delta=1$ and go from $N_{\text{steps}}=0$ to $N_{\text{steps}}=6$. As a result, only finitely-spaced values of $U$ can be chosen. Here, smaller values of $U$ involve fewer gates and will be less affected by experimental errors.

Method II: We fix $N_{\text{steps}}=5$, and can hence sample any value of $U$ up to $N_{\text{steps}}$. In this case, $\tau=N_{\text{steps}}\delta/U$. Here, the same number of gates are performed at every value of $U$ and so the magnitude of the experimental error should be similar at every point.

Based on simulations (see Supplementary Material), we choose the parameters that seem to most closely follow the results from exact diagonalization. For method I, $\delta=0.1$ and $\tau=0.1$, while for method II, $\delta=0.25$ and $\tau=1.25/U$.

\begin{figure}[t]
\includegraphics[width=0.9\columnwidth]{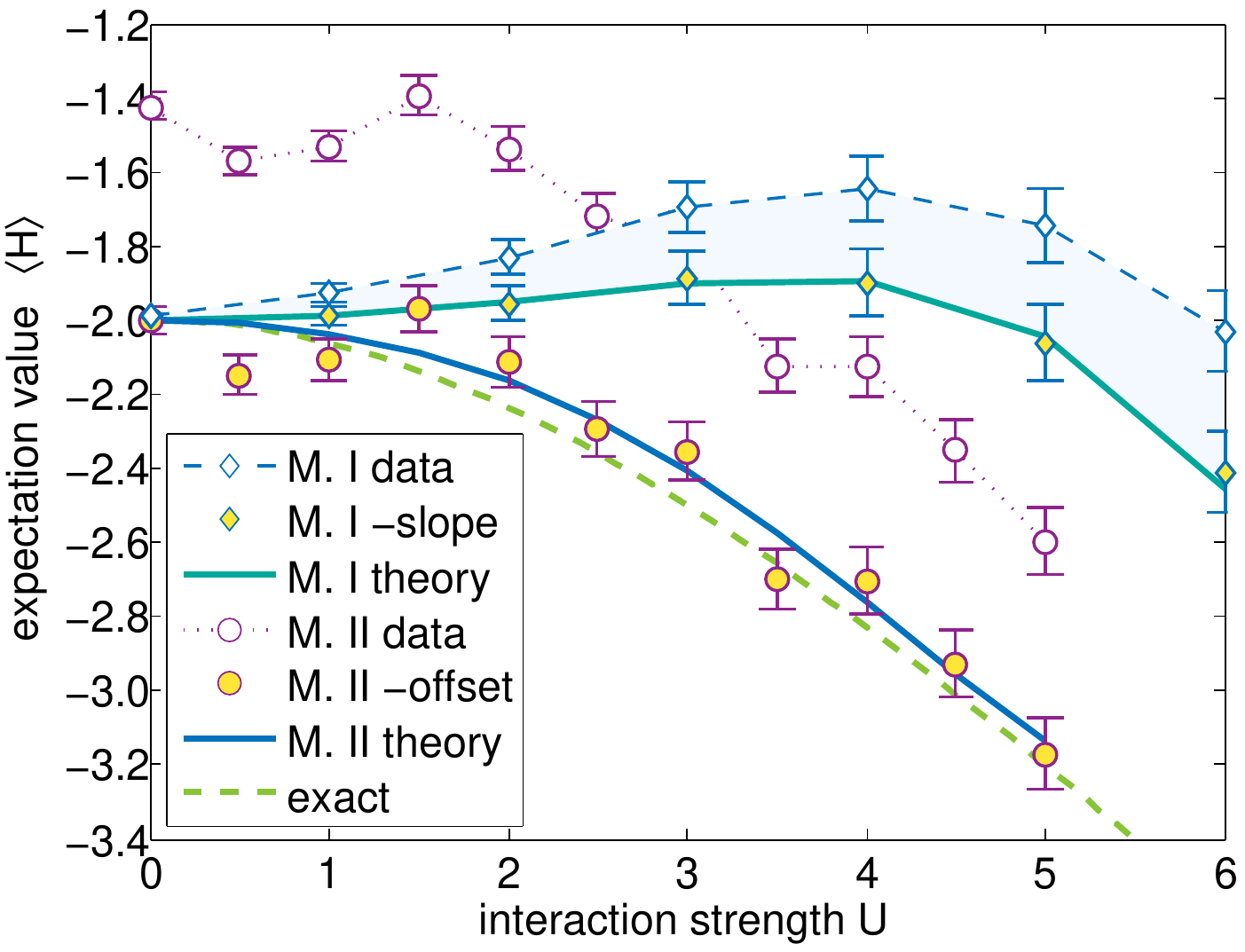}
\caption{Expectation value of the Hamiltonian $\langle H \rangle$ of the Fermi-Hubbard model after adiabatic evolution, to the values of $U$ given by the abscissa. Method I (M. I, open diamonds) sees an increase in experimental error with $U$ (shaded area) since $N_{\textrm{steps}}$ increases with $U$. Subtracting a straight line of slope $0.063$ best matches the data points to the theory curve (filled diamonds). For method II (M. II, open circles), $N_{\textrm{steps}}$=5 is constant and the experimental error in $\langle H \rangle$ is seen as an offset. Even for large systems, this can be measured for the integrable case $U=0$ and then subtracted (filled circles). The corrected results (measured offset $0.58$) follow the theory curve well. Method II is a better match to the exact solution.}
\label{fig:adiabaticevolution}
\end{figure}

We first implement the digitized adiabatic evolution by itself for both methods in a two-qubit experiment on our system. We prepare the qubits along the $x$-axis of the Bloch sphere, in the ground state of the non-interacting Hamiltonian ($U=0$) using Hadamard gates (see Fig. \ref{fig:circuit}). Then we evolve the state in steps from $U=0$ to $U=6$ for method I and $U=5$ for method II. Finally we measure along the $z$-axis and separately (with the help of additional Hadamard gates) along the $x$-axis to calculate the expectation value $\langle H \rangle$ from $\langle X_1 \rangle$,  $\langle X_2 \rangle$, and $\langle Z_1 Z_2 \rangle$. The results are shown in figure \ref{fig:adiabaticevolution}. We see that method II shows an offset from the ideal value. If we subtract the value measured for $U=0$ (no evolution) from all data points, they match the theoretical expectation closely. For method I, the number of gates and hence the error incurred grows with $U$. Subtracting a straight line of slope $0.063$ models this increase well. In a larger system, where the dynamics are unknown, this correction cannot be easily determined. The offset seen in method II, however, corresponds to the error in the eigenvalue of a non-interacting and hence easily integrable system, which can be more generally applied. 


To measure $R_2$, we follow the technique outlined in \cite{Johri17}, which requires two copies of the state $|\Psi\rangle=\sum_{i,j}c_{ij}|a_i\rangle|b_j\rangle$. $R_2$ is given by the expectation value of the Swap operator on subspace $A$, 
\begin{equation}
R_2=\langle\Psi|\langle\Psi|\text{Swap}_A|\Psi\rangle|\Psi\rangle,
\end{equation} 
where the operator $\text{Swap}_A$ acts as follows:
\begin{align}
	\text{Swap}_A |\Psi\rangle|\Psi\rangle =\sum_{i,j}\sum_{i',j'}c_{ij}c_{i'j'}|a_{i'}\rangle|b_j\rangle |a_i\rangle|b_{j'}\rangle .
	\label{eq:cswap}
\end{align}
To extract $R_2$ experimentally, we apply the Swap-gate to the subsystems $A$ of two copies of the adiabatically evolved state, conditional on the state of an ancilla qubit. The ancilla qubit is prepared and measured in the X-basis by applying a Hadamard gate before and after the C-Swap gate (see figure \ref{fig:circuit}). Repeating the measurement and averaging allows us to determine the probability $P_a$ to find the ancilla qubit in state $|0\rangle$ or $|1\rangle$ from which $R_2$ is calculated as $R_2=P_a(0)-P_a(1)$.

\begin{figure}[t]
\centering
\includegraphics[width=0.78\columnwidth]{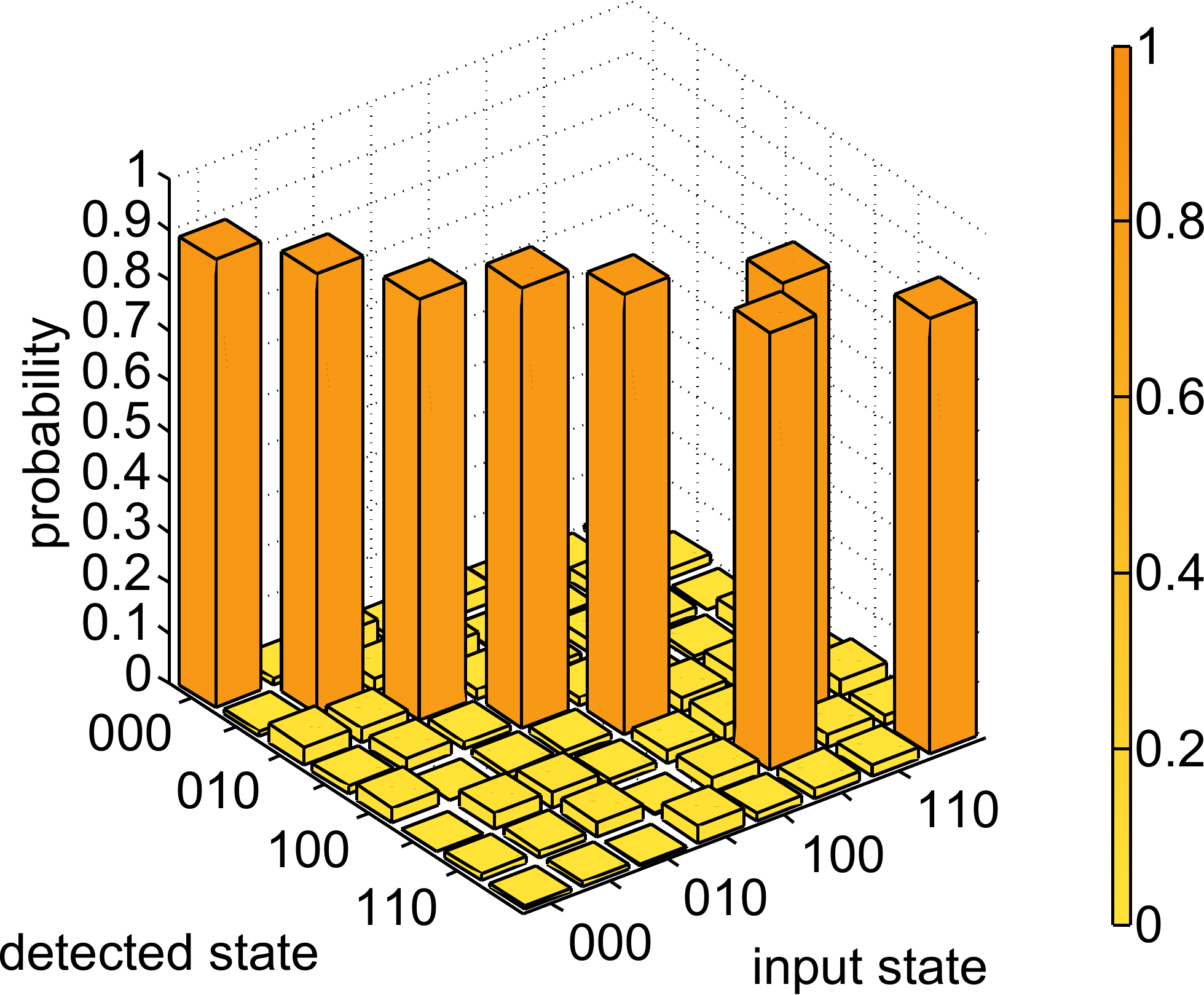}
\caption{Implementation of the C-Swap or Fredkin gate. The plot shows results for all eight input states, reproducing the state transfer matrix (eqn. \ref{eq:fredkin}) with an average success probability of $86.8(3)\%$, while the state of the control qubit is correct with $94.0(2)\%$ probability. The results have been corrected for $\sim 1\%$ state-preparation and measurement error.}
\label{fig:fredkin}
\end{figure}

The C-Swap or Fredkin gate \cite{Fredkin1982} has been experimentally implemented in NMR \cite{Du2006} and photonic systems \cite{Patel16, Ono16}. Our work is the first implementation of a C-Swap gate with trapped ions. Its state transfer matrix is 
\begin{equation}
U_{\textrm{C\textrm{-}Swap}}=\left(
\begin{smallmatrix}
1 & 0 & 0 & 0 & 0 & 0 & 0 & 0 \\
0 & 1 & 0 & 0 & 0 & 0 & 0 & 0 \\
0 & 0 & 1 & 0 & 0 & 0 & 0 & 0 \\
0 & 0 & 0 & 1 & 0 & 0 & 0 & 0 \\
0 & 0 & 0 & 0 & 1 & 0 & 0 & 0 \\
0 & 0 & 0 & 0 & 0 & 0 & 1 & 0 \\
0 & 0 & 0 & 0 & 0 & 1 & 0 & 0 \\
0 & 0 & 0 & 0 & 0 & 0 & 0 & 1 \\
\end{smallmatrix}
\right).
\label{eq:fredkin}
\end{equation}
We realize this gate in our system by programming the quantum compiler to break it down into gates from our native library \cite{Debnath16, LinkePNAS2017}. It requires seven entangling gates and fourteen single-qubit rotations. A circuit diagram detailing its modular implementation is shown in the Supplementary Material. We test this gate by applying each logical input state and recording the output state probabilities. The results are shown in figure \ref{fig:fredkin}. Compared to the ideal state transfer matrix shown in equation \ref{eq:fredkin}, the average success probability of this gate is $86.8(3)\%$. The control qubit, on which the measurement of $R_2$ hinges, is found to be in the correct state with $94.0(2)\%$ probability.


With all elements of the circuit in place, we now implement the entire algorithm to adiabatically evolve two copies of the two-qubit Fermi-Hubbard system and measure the second Renyi entropy using the top qubit as the ancilla (see Fig. \ref{fig:circuit}). The results are shown as open symbols for method I and II in figure \ref{fig:compresults}(a) and (b), respectively. Both figures include a curve showing the theoretically expected values for the chosen Trotter step ($\delta$) and evolution time ($\tau$), as well as the exact solution for comparison. The exact solution shows $R_2$ monotonically decreasing with increasing interaction, implying increasing entanglement between the qubits. For method I the curve starts just below $0.9$. At this point, no adiabatic evolution is applied, and this value is expected from the performance of the C-Swap gate. The deviation from the theoretically expected curve increases as more evolution steps are taken. For method II, $27$ entangling gates have to be performed regardless of the value of $U$ and we observe a systematically lowered value of $R_2$. It is clear that the probability distribution of the ancilla qubit is determined by both the dynamics of the model and the errors in the physical gates.


To distinguish between these two phenomena, we develop a method to detect erroneous runs using the additional information available in the four data qubits representing the two systems. 
We know that the eigenvalues of the Swap operator are $\pm 1$ with the corresponding eigenstates being even and odd functions of the qubits being swapped. For any operator $V$, with eigenvalues $\lambda_m$ and eigenvectors $|m\rangle$ such that $V|m\rangle=\lambda_m|m\rangle$, and a state $|\Phi\rangle=d_m|m\rangle$, the expectation value $\langle\Phi|V|\Phi\rangle$ can be obtained by applying $V$ conditional on the state of an ancilla qubit. The ancilla is prepared along $x$ before the controlled-V operation and measured in the X-basis thereafter.
Just before measurement, the qubits are in the state:
\begin{align}
|0\rangle\sum_m(1+\lambda_m)d_m|m\rangle+|1\rangle\sum_m(1-\lambda_m)d_m|m\rangle.
\end{align}
This is essentially the circuit used to measure the expectation value of the $\text{Swap}_A$ operator to determine the Renyi entropy. 
Therefore we can make the observation that the probability of $|1\rangle |a_i\rangle|b_j\rangle |a_{i'}\rangle|b_{j'}\rangle$ is 0 if $i'=i$ or if $j'=j$. This implies that twelve of the $32$ possible output states of the $5$-qubit register should have zero-weight when it comes to evaluating $R_2$ (see Supplementary Material). We re-analyze the data after discarding such outcomes and find the values given by the filled symbols in figure \ref{fig:compresults}. The data points now follow the theoretical curves, showing that the method succeeds in substantially reducing experimental errors. This technique is general since it only depends on the symmetry of the C-Swap operation and is independent of the evolved state or model Hamiltonian under investigation. The yield, or fraction of data runs that are kept under this method, is $\sim84\%$ for method II, and drops with $U$ from $94\%$ to $83\%$ for method I (see Supplementary Material). For the deeper circuits, the technique is not able to discard all errors, however, as can be seen in figure \ref{fig:compresults}(a) at $U=6$.

In summary, we have demonstrated a complete chain of steps (i.e. the `full quantum computing stack') to simulate a model Hamiltonian on a quantum computer and measure bipartite entanglement. Each step is scalable to a larger system of qubits on the trapped-ion hardware platform. The technique can also be generalized to an arbitrary Hamiltonian and implemented on different quantum computing architectures. 

\begin{acknowledgments}
This work was supported by the IARPA LogiQ program, the ARO Atomic Physics program, the ARO MURI on Modular Quantum Circuits, the AFOSR MURI program on Optimal Quantum Measurements, and the NSF Physics Frontier Center at JQI.
\end{acknowledgments}


%

\newpage
\clearpage
\section{Supplementary material}

\subsection{Experimental system}

We perform the experiment on a quantum computer consisting of a chain of five $^{171}$Yb$^+$ ions confined in a Paul trap and laser cooled near the motional ground state. The hyperfine-split $^2S_{1/2}$ ground level with an energy difference of $12.642821\:$GHz provides a pair of qubit states, which are magnetic field independent to first order. The typicial coherence time of this so-called ``atomic clock'' qubit is $~0.5\:$s, which can be straightforwardly extended by reducing magnetic field noise. Optical pumping is used to initialize the state of all ions, and the final states are measured collectively via state-dependent fluorescence detection \cite{Olmschenk07}. Each ion is mapped to a distinct channel of a photomultiplier tube (PMT) array. The average state detection fidelity is $99.4(1)\%$ for a single qubit, while a $5$-qubit state is typically read out with $95.7(1)\%$ average fidelity, limited by channel-to-channel crosstalk. These state detection and measurement (SPAM) errors are characterized in detail by measuring the state-to-state error matrix. For averaged data such as the one shown in this work, they are straightforwardly corrected by re-normalizing the averaged state vector by the inverse of this matrix. Quantum operations are achieved by applying two Raman beams from a single $355\:$nm mode-locked laser, which form beat notes near the qubit frequency. The first Raman beam is a global beam applied to the entire chain, while the second is split into individual addressing beams, each of which can be switched independently to target any single qubit \cite{Debnath16}. Single qubit gates are generated by driving resonant Rabi rotations (R-gates) of defined phase, amplitude, and duration. Two-qubit gates (so-called XX-gates) are realized by illuminating two ions with beat-note frequencies near the motional sidebands and creating an effective spin-spin (Ising) interaction via transient entanglement between the state of two ions and all modes of motion \cite{Molmer99, Solano99, Milburn00}. To ensure that the motion is left disentangled from the qubit states at the end of the interaction, we employ a pulse shaping scheme by modulating the amplitude of the global beam \cite{Zhu06eur,Choi14}. The signal to drive each ion is generated by an individual Arbitrary Waveform Generator (AWG) which allows us to efficiently apply single-qubit Z-rotations as classical phase advances.

\subsection{Error and Resource Analysis}

\begin{figure}
	\includegraphics[width=\columnwidth]{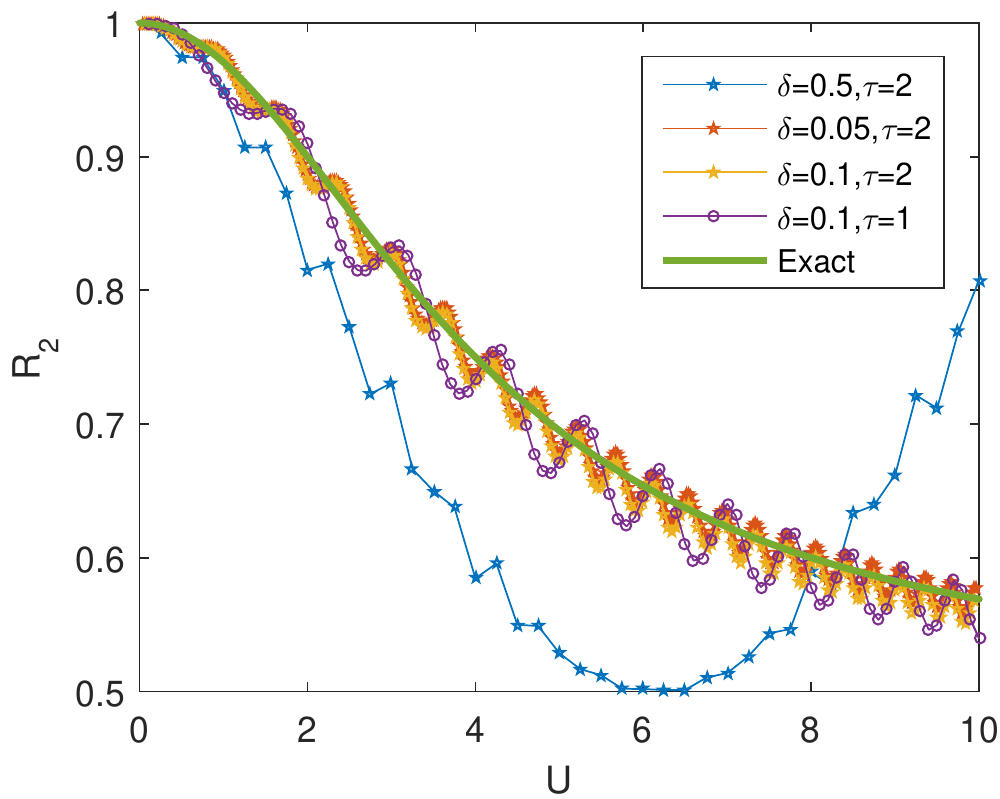}
	\caption{The value of $R_2$ output from the quantum algorithm compared to the exact value at different values of the adiabatic evolution time $\tau$ and Trotter step size $\delta$ for the first-order Trotter approximation.}
	\label{fig:r2}
\end{figure}

Fig. \ref{fig:r2} shows the effect of varying the Trotter step size $\delta$ and the adiabatic evolution time $\tau$ on the accuracy of the measured $R_2$ as compared to the exact value between $U=0$ and $U=10$.  We only consider the first order Trotter approximation because with only two non-commuting terms in the Hamiltonian, the first and second order approximations are asymptotically equivalent. We see that increasing $\tau$ increases the amplitude of the oscillations around the exact result, whereas increasing $\delta$ leads to an increase in a constant offset from the exact value. When $\delta$ becomes too large (compared to the largest eigenvalue of the Hamiltonian), the Trotter approximation breaks down.

\begin{figure*}
	\includegraphics[width=\textwidth]{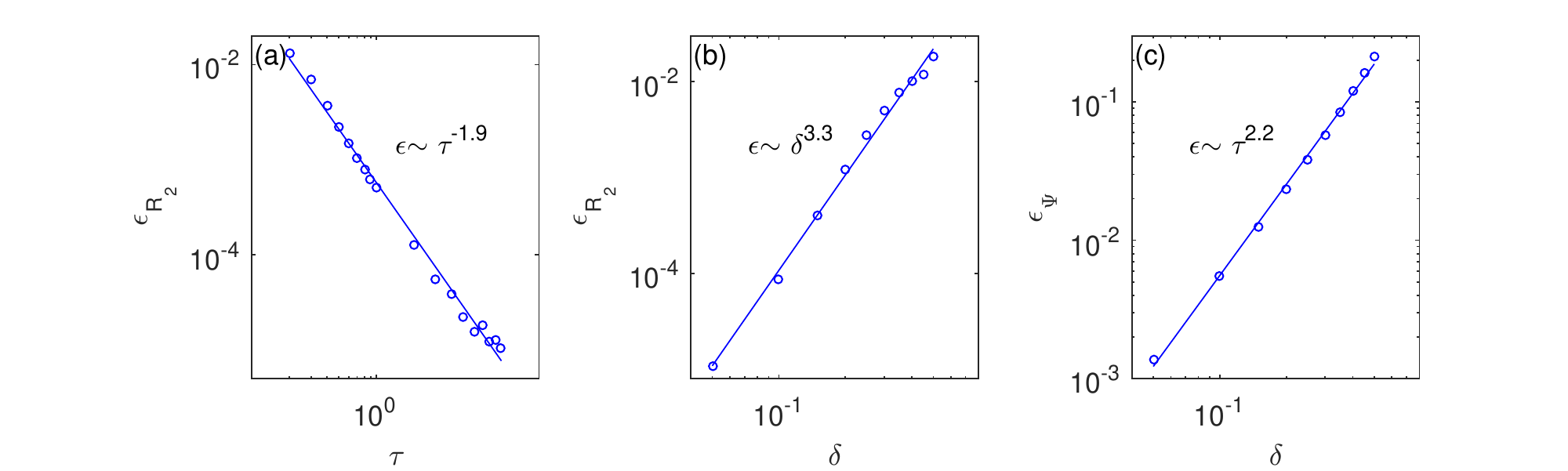}
	\caption{(a) The error in $R_2$, $\epsilon_{R_2}$ (defined in the text) as a function of evolution time $\tau$ at fixed $\delta=0.05$ fits well to a line of slope -1.9. (b) $\epsilon_{R_2}$ as a function of Trotter step size $\delta$ at fixed $\tau=10$ fits well to lines of slope 3.3. (c) The error in the wavefunction $\epsilon_{\Psi}$ (defined in the text) at fixed $\tau=10$ as a function of Trotter step size $\delta$ fits well to a line of slope $2.2$.}
	\label{fig:r2_error}
\end{figure*}

Fig. \ref{fig:r2_error} shows how the error from the first-order Trotter approximation varies with different parameters. The error in $R_2$ is defined as the chi-squared goodness of fit of the output of the quantum algorithm to the actual data between U=0 to U=10, that is 
\begin{align}
\epsilon_{R_2}=\overline{\bigg(\frac{(R_2^{\text{exact}}-R_2^{\text{sim}})^2}{R_2^{\text{exact}}}\bigg)},
\end{align} 
where the average is over $U=0$ to $U=10$ measured at intervals of $U=0.1$. As expected from the adiabatic evolution theorem, $\epsilon_{R_2}\sim\tau^{-2}$, which is the behavior seen in Fig. \ref{fig:r2_error}(a). 
 
The error in the wavefunction, defined as the projection of the wavefunction orthogonal to the ground state, averaged between U=0 to U=10, is given by
\begin{align}
\epsilon_{\Psi}=\overline{1-\langle\Psi^{\text{exact}}|\Psi^{\text{sim}}\rangle^2}.
\end{align}
The data show that the error in the wavefunction and the error in the actual quantity of interest can scale differently.

\begin{figure*}
	\includegraphics[width=\textwidth]{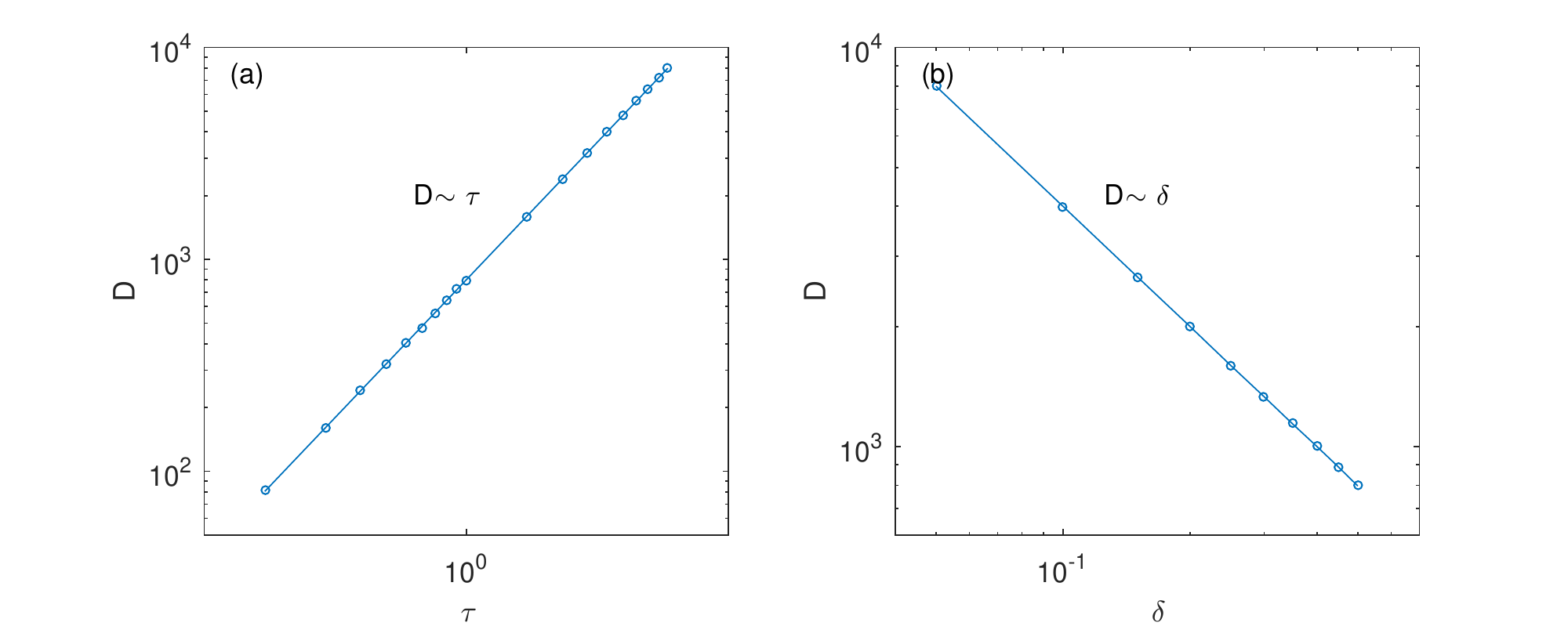}
	\caption{In first Trotter approximation, (a) the parallel circuit depth $D$ as a function of evolution time $\tau$ at fixed $\delta=0.05$ fits well to lines of slope $1$,  and (b) $D$ as a function of Trotter step size $\delta$ at fixed $\tau=10$ fits well to lines of slope $-1$. Here $D$ is defined as the circuit depth for adiabatic evolution to $U=10$.}
	\label{fig:depth}
\end{figure*}

Fig. \ref{fig:depth} shows the circuit depth as a function of $\delta$ and $\tau$ scales linearly with each, which is to be expected from the first-order Trotter approximation and adiabatic evolution.

\subsection{Parameter choices for the experiment}

\begin{figure*}
	\includegraphics[width=0.9\textwidth]{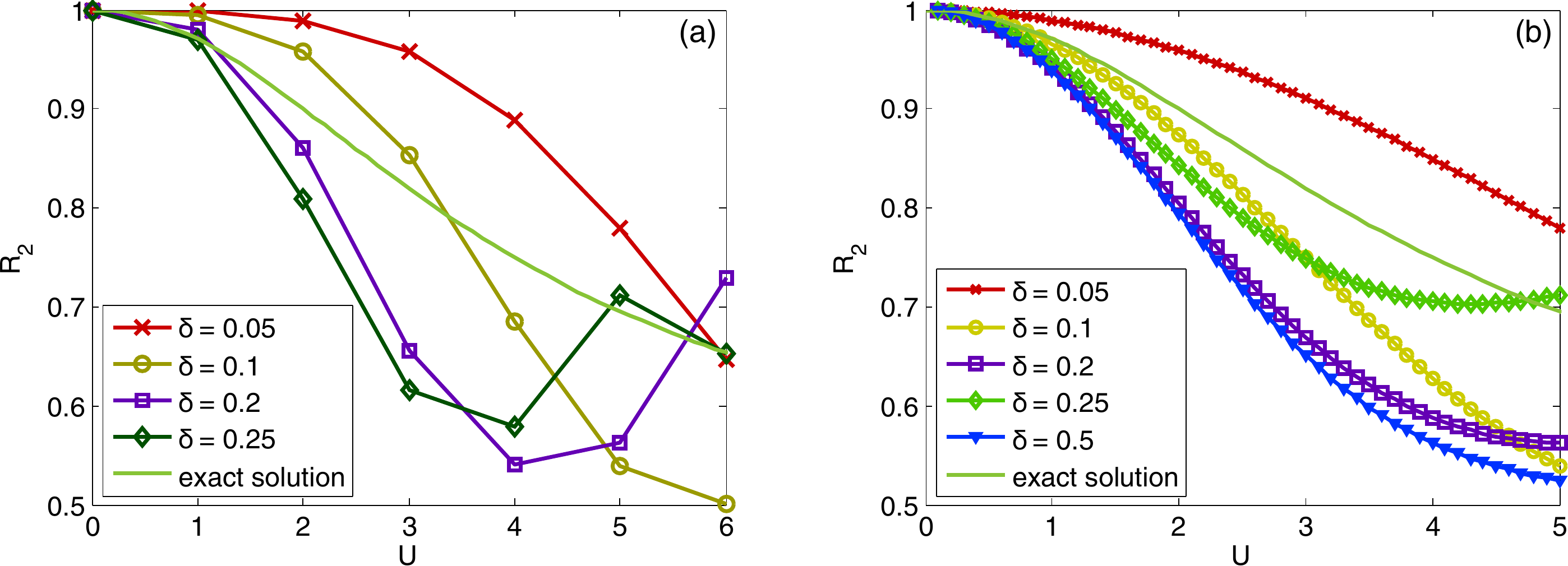}
	\caption{The value of $R_2$ resulting from the quantum algorithm compared to the exact value for different values of the adiabatic evolution time $\tau$ and Trotter step size $\delta$ for (a) method I with fixed $\tau/\delta=1$, and (b) method II with fixed $N_{\text{steps}}=5$.}
	\label{fig:parameters}
\end{figure*}

Figure \ref{fig:parameters} shows theoretical calculations of the outcome of the $R_2$ measurement for various parameters that satisfy the experimental restriction $N_{\text{steps}}\leq 6$. For the experiment, the following sets were chosen: For method I, $\delta=\tau=0.1$, and for method II, $\delta=0.25$ and $\tau=1.25/U$.

\subsection{The C-Swap or Fredkin gate}

\begin{figure*}[htb]
\hspace{1em}\Qcircuit @C=1em @R=0.7em @!R {
\lstick{1} & \ctrl{2} 	& \qw & & &
\qw  						&\qw									&\gate{R_x(\gamma\frac{\pi}{2})} 	&\gate{R_z(-\frac{\pi}{2})}						& \qw    								& \gate{XX_{13}(\gamma\frac{\pi}{8})} & \qw & \rstick{\cdots} \\
\lstick{2} & \qswap 	& \qw &\push{\rule{.3em}{0em}=\rule{.3em}{0em}} & & 
\qw 						&\multigate{1}{XX_{23}(\beta\frac{\pi}{4})} 	&\gate{R_z(-\beta\frac{\pi}{2})}		&\gate{R_x(\frac{\pi}{4}+(1-\beta)\frac{\pi}{4})} 				& \multigate{1}{XX_{23}(\beta\frac{\pi}{8})} 	& \qw\qwx & \qw & \rstick{\cdots}\\
\lstick{3} & \qswap  	&\qw & & &  
\gate{R_y(\beta\frac{\pi}{2})} 	&\ghost{XX_{23}(\beta\frac{\pi}{4})}			& \gate{R_z(-\frac{\pi}{2})}  		&\gate{R_x(-\beta\frac{\pi}{2}+\frac{\pi}{4})} 		& \ghost{XX_{23}(\beta\frac{\pi}{8})}	 		& \gate{XX_{13}(\gamma\frac{\pi}{8})} \qwx & \qw & \rstick{\cdots}
}
\vspace{2em}
\hfill\Qcircuit @C=1em @R=0.7em @!R {
& & & & & & & \lstick{\cdots} 	&\gate{R(-\frac{2\pi}{3}, (\frac{\gamma+1}{2})\pi - P)} 	& \multigate{1}{XX_{12}(\alpha\frac{\pi}{4})}	& \qw & \gate{R(-\alpha\beta\gamma\frac{2\pi}{3}, (\frac{\alpha\beta+1}{2})\pi - \alpha\beta\gamma P)}	& \gate{XX_{13}(\gamma\frac{\pi}{8})} & \qw & \rstick{\cdots}\\
& & & & & & & \lstick{\cdots} 	& \qw 												& \ghost{XX_{12}(\alpha\frac{\pi}{4})}	& \qw & \qw 																		& \qw\qwx & \qw & \rstick{\cdots}\\
& & & & & & & \lstick{\cdots}	&\qw  												&\qw								& \qw	& \qw 																		& \gate{XX_{13}(\gamma\frac{\pi}{8})}\qwx & \qw & \rstick{\cdots}
}
\vspace{2em}
\hfill \Qcircuit @C=1em @R=0.7em @!R {
& & & & \lstick{\cdots}	& \gate{R(\pi,-\alpha\beta\gamma\frac{\pi}{4})} 	& \multigate{1}{XX_{12}(\alpha\frac{\pi}{4})} 	& \qw & \qw & \qw & \qw & \qw \\
& & & & \lstick{\cdots} 	& \qw 									& \ghost{XX_{12}(\alpha\frac{\pi}{4})} 		& \gate{R_z(-\beta\frac{\pi}{2})}   & \multigate{1}{XX_{23}(\beta\frac{\pi}{4})}	& \qw & \qw & \qw \\
& & & & \lstick{\cdots}  & \qw  									& \qw  								& \gate{R_y(\beta\frac{\pi}{2})}  	 & \ghost{XX_{23}(\beta\frac{\pi}{4})}			& \gate{R_y(-\beta\frac{\pi}{2})} & \gate{R_z(-\frac{\pi}{2})} & \qw
}
\caption{Controlled-Swap (or Fredkin) gate implementation using entangling $XX(\chi)$ gates, and single qubit rotations $R_x(\theta)$, $R_y(\theta)$, and $R(\theta,\phi)$. The parameters are $\alpha = sgn(\chi_{12})$, $\beta = sgn(\chi_{23})$, $\gamma = sgn(\chi_{13})$, and $P = \arcsin{\sqrt{\frac{2}{3}}}$}.
\label{fig:Fredkincircuit}
\end{figure*}
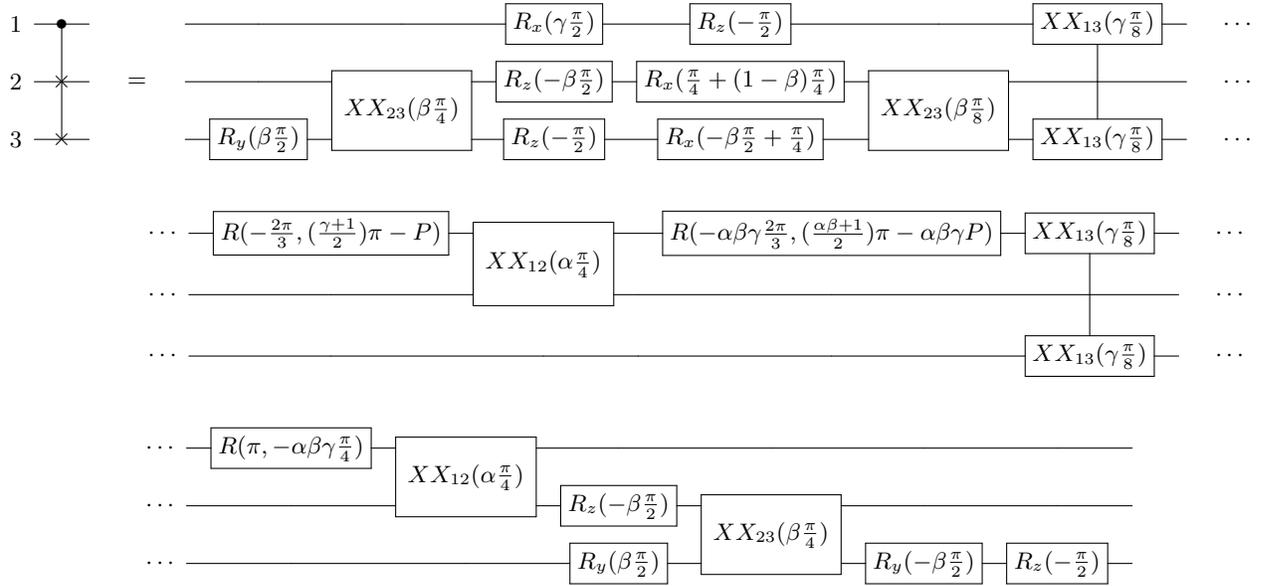

The compiler breaks down the Fredkin (C-Swap) gate into native R- and XX-gates as given by the circuit in figure \ref{fig:Fredkincircuit}.


\subsection{Error reduction}

\begin{figure}
	\includegraphics[width=\columnwidth]{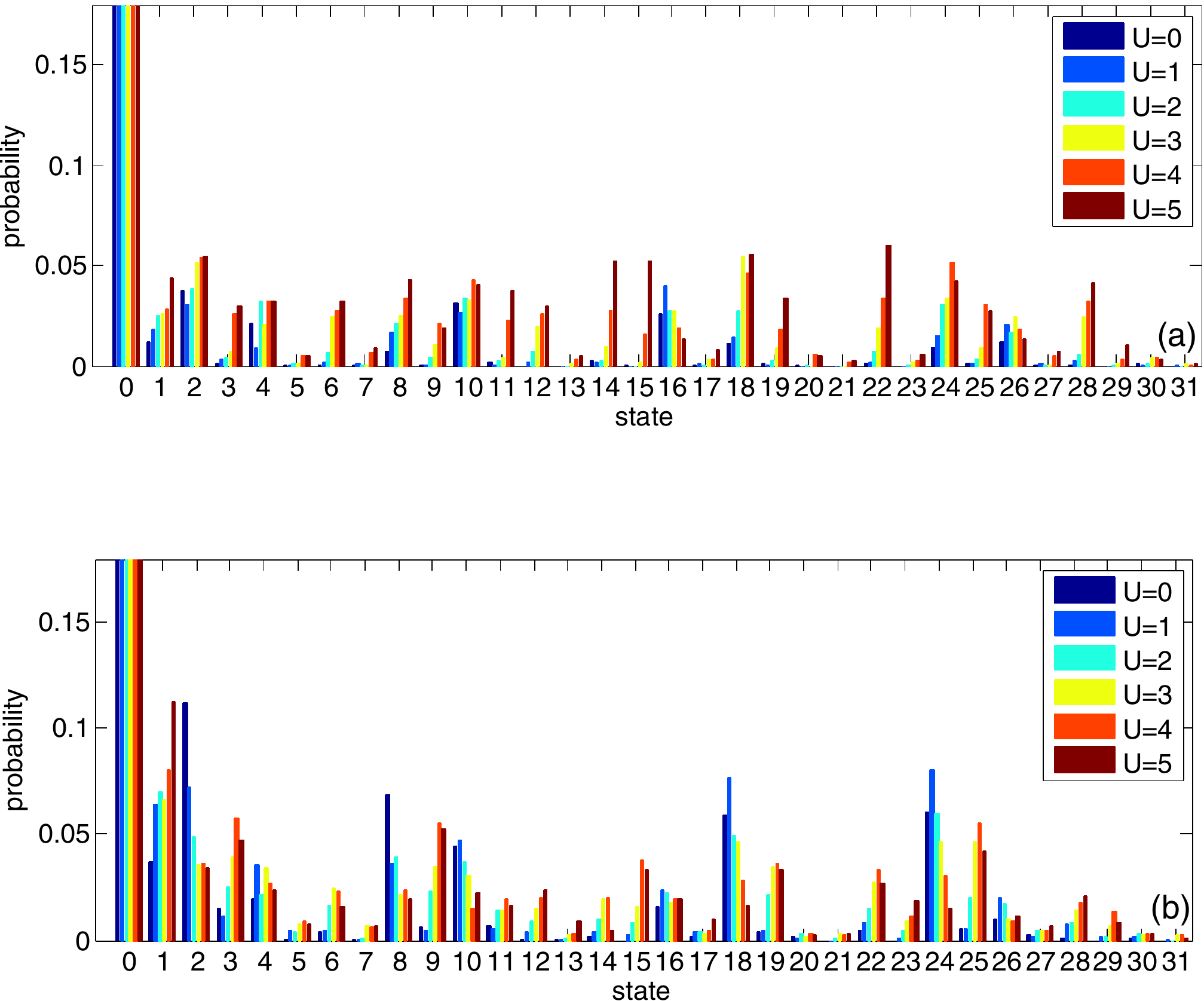}
	\caption{The original output of the circuit shown in figure \ref{fig:circuit} averaged over $2000$-$2500$ runs, corrected for SPAM errors. Method I is shown in (a) and method II in (b).}
	\label{fig:original}
\end{figure}

\begin{figure}
	\includegraphics[width=\columnwidth]{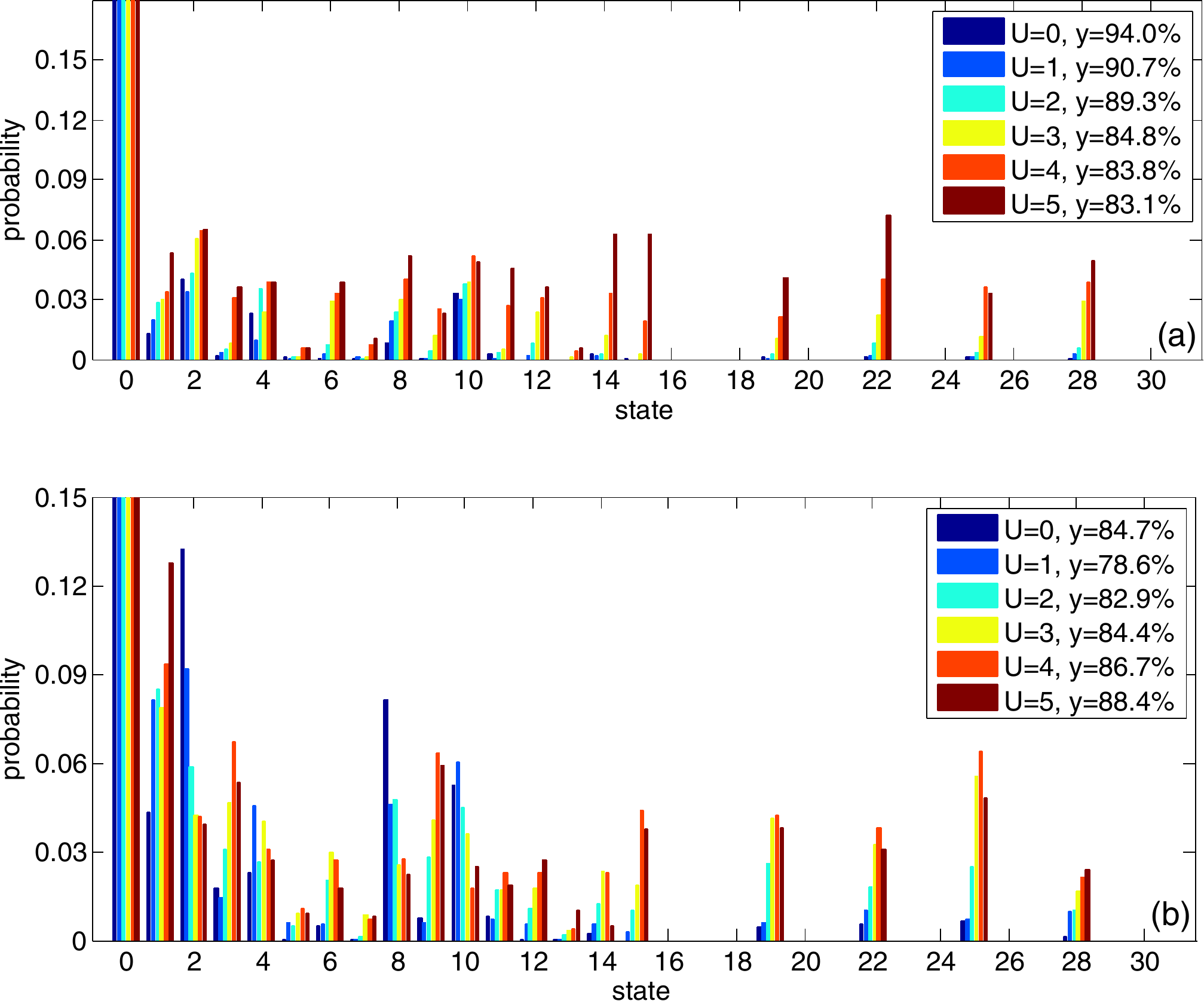}
	\caption{The output of the circuit shown in figure \ref{fig:circuit} averaged over $2000$-$2500$ runs, corrected for SPAM errors and after discarding in post-selection any runs that resulted in the states given in the text. Method I is shown in (a) and method II in (b). The legend gives the yields, i.e. the fraction of experimental runs that was kept.}
	\label{fig:post}
\end{figure}
Using the convention that the top-most qubit in the five-qubit output state after the circuit shown in figure \ref{fig:circuit} is the most significant bit in the binary representation of the computational basis, we can assign each state from $|00000\rangle$ to $|11111\rangle$ a decimal value from $0$ to $31$. The observation about the symmetry of the state after the C-Swap gate implies that the states numbered 16, 17, 18, 20, 21, 23, 24, 26, 27, 29, 30, and 31 should have zero-weight at the end of the circuit. This can be used to post-select the results by discarding runs that result in these outcomes. Figure \ref{fig:post} shows the results before and after this step. It also gives the yields associated with each experiment. We see that we discard less than $20\%$ of the data and achieve an improvement in the results of over $40\%$ (see Fig. \ref{fig:compresults}).

\end{document}